\documentclass[preprint,12pt,3p,times,sort&compress]{elsarticle}

\usepackage{amssymb}
\usepackage{amsmath}
\usepackage[T1]{fontenc}
\usepackage{lmodern}

\usepackage{bbm}
\usepackage{epsfig}
\usepackage{epstopdf}
\usepackage{color}

\usepackage{hyperref}

\definecolor{mycolor}{rgb}{0,0,0}

\journal{Physica A}

\begin{document}

\begin{frontmatter}



\title{Impact of memory on opinion dynamics}


\author{Arkadiusz J\k{e}drzejewski}
\ead{arkadiusz.jedrzejewski@pwr.edu.pl}

\author{Katarzyna Sznajd-Weron}
\ead{katarzyna.weron@pwr.edu.pl}
\address{Department of Theoretical Physics, Faculty of Fundamental Problems of Technology, Wroc\l{}aw University of Science and Technology, Wroc\l{}aw, Poland}

\begin{abstract}
We investigate an agent-based model of opinion dynamics with two types of social response: conformity and  independence. Conformity is introduced to the model analogously as in the Sznajd model or $q$-voter model, which means that only unanimous group exerts peer pressure on individuals. The novelty, in relation to previous versions of the $q$-voter model, is memory possessed by each agent and external noise $T$, which plays the role of social temperature. Each agent has its own memories of past experiences related to the social costs and benefits of being independent or conformist. If an agent was awarded in past more for being independent, it will have a greater tendency to be independent than conformist and vice versa. We will show that depending on the social temperature $T$ the system spontaneously organizes into one of two regimes.
Below a certain critical social temperature $T_c$, all agents in the society acquire personal traits (so called \textit{person state}). Some of them become permanent conformists and others start to behave forever independently. This means that initially homogeneous population becomes heterogeneous, and agents respond differently to social influence. For $T>T_c$, all agents with equal probabilities behave independently or conform to peer pressure (so called \textit{situation state}). This regime change between person and situation state, which reminds the idea of an annealed vs. quenched disorder, affects also public opinion. Particularly interesting results are obtained for individualistic societies, in which public opinion is non-monotonic function of $T$, which means that there is an optimal social temperature for which an agreement in the society is the highest. \\\\
Accepted Manuscript of A. J\k{e}drzejewski, K. Sznajd-Weron, \href{https://doi.org/10.1016/j.physa.2018.03.077}{Physica A \textbf{505}, 306 (2018)}.\\
DOI: 10.1016/j.physa.2018.03.077\\
\textcopyright\:2018. This manuscript version is made available under the CC-BY-NC-ND 4.0 license\\
\url{https://creativecommons.org/licenses/by-nc-nd/4.0/}

\end{abstract}


\end{frontmatter}


\section{Introduction}
\label{sec:intro}
As written by Daniel Kahneman in the article based on his Nobel Prize lecture: ``\textit{A theory of choice that completely ignores feelings such as the pain of losses and the regret of mistakes is not just descriptively unrealistic. It also leads to prescriptions that do not maximize the utility of outcomes as they are actually experienced (...).}'' \cite{Kah:03}. This means that individual's memories influence judgments and choices that people make. Yet, most of opinion dynamics models, which can be treated as a zero-level approach to various more complex social processes like political voting, marketing choices, diffusion of innovation, etc., initially did not include memory on the microscopic level. This applies in particular to binary-opinion models, such as: the voter model (VM) \cite{Cli:Sud:73,Sen:Bik:13}, models of social impact \cite{Now:Sza:Lat:90,Hol:Kac:Sch:00}, the Galam (majority) model \cite{Gal:86,Gal:90,Gal:12}, the Sznajd model \cite{Szn:Szn:00,Szn:05a}, the threshold model \cite{Wat:02}, or the $q$-voter model \cite{Cas:Mun:Pas:09}. 

Due to our knowledge, the idea of memory was introduced for the first time into the voter model by Dall'Asta and Castellano to reduce the noise of VM, which resulted in the appearance of an effective surface tension \cite{Dal:Cas:07}. They endowed each $i$-th voter with a pair of counters:  $C^+_i$ for being in a state '$+$' and $C^-_i$ for being in a state '$-$'. At each time step the corresponding counter was updated, and a voter could change its state only when the counter reached a given threshold value. The similar idea was used later in the model describing consumer decisions regarding switching to dynamic electricity tariffs \cite{Kow:etal:14}. However, in the latter paper only one counter was introduced to measure for how many steps the same state (opinion) is kept, which reflected the assumption that decision is based on the unanimity of past opinions. Since 2007, the concept of memory has been incorporated into the voter model in several ways \cite{Sen:Bik:13}. However, all proposed approaches included one or another waiting time for the opinion change \cite{Sta:Tes:Sch:08,Xio:Liu:11,Tak:Mas:11}. For example, Stark \textit{et al.} studied the effect of a memory-dependent transition rate in the voter model assuming that the flip rate decreases with the time the voter has been in its current state \cite{Sta:Tes:Sch:08}. Xiong \textit{et al.} have assumed that an individual inclination increases with the number of times the voter has held its most frequent opinion in the past interactions \cite{Xio:Liu:11}. Different idea of memory has been proposed by Takaguchi and Masuda, who endowed links, instead of voters, between the nodes with a random variable, which represented the time until the initial update event occurs on this link \cite{Tak:Mas:11}. Zhong \textit{et al.} have introduced a generalized voter model with time-decaying rate of the influence of peer pressure, which incorporates a multilayer network and memory of past influences \cite{Zho:etal:16}.

Recently, another idea of memory has been introduced into the voter model by Woolcock \textit{et al.} \cite{Woo:etal:17}. They studied a heterogeneous voter model in which each agent was endowed with the fitness parameter. During pairwise interactions agents' fitnesses are compared, and with probability $p$ the agent with the lower fitness adopts the opinion of the one with the higher value, whereas with complementary probability the opposite happens. The winning agent (the one that keeps its opinion) increases its fitness. This idea reminds a lot of the Bonabeau model, in which agents also fight when they meet. Each fight influences the agent's power: a winner becomes stronger, and a looser becomes weaker, what alters the probabilities of winning their future fights \cite{Bon:The:Den:95,Mal:Sta:Kul:06}. 

In this work, we propose to introduce memory into opinion dynamics in yet another way inspired by a paper on random walkers with extreme value memory \cite{Har:15}. We will study the $q$-voter model, which has occurred to be particularly interesting from theoretical and applicative  point of view \cite{Cas:Mun:Pas:09,Mor:etal:13,Tim:Pra:14,Jav:Squ:15,Mob:15,Tim:Gal:15,Chm:Szn:15,Sie:Szw:Wer:16,Mel:Mob:Zia:16, Kru:Szw:Wer:17,Jed:17,Jed:Szn:17}. We will focus on the $q$-voter model with independence, introduced in \cite{Nyc:Szn:Cis:12}. This version of the model occurred to be particularly useful in modeling diffusion of eco-innovations \cite{Kow:etal:14,Kow:etal:16,Byr:etal:16,Kow:17}. Until now the level of independence $p$ was an external parameter of the model, and two different approaches have been used: person and situation-oriented \cite{Szn:Szw:Wer:14,Jed:Szn:17}. Within the situation-oriented approach the system is homogeneous, and   each of $N$ agents can behave independently with probability $p$ and conform to the group with complementary probability $1-p$. On the other hand, within the person-oriented approach $pN$ agents are permanently independent, and $(1-p)N$ of them always conform to the source of influence. In this paper we will show that without assuming a priori person-oriented approach and setting particular value of parameter $p$, we can obtain two regimes (situation or person state) depending on the external noise $T$, which can be interpreted as the social temperature \cite{Kac:Hol:96,Bah:Pas:98}. For low $T$, the system will self-organize from the situation state, in which the population is homogeneous, to the state in which agents acquire personal traits. This self-organization will appear as a result of bad and good memories related to the type of social response. 

There are many social benefits of conformity, like sense of security, fraternity, or convenience. It is easier to cooperate with people that follow social norms. On the other hand, we often think of conformity as a bad thing, meaning that people who conform are weak and dependent. Especially in western individualistic cultures, the word ``conformity'' tends to carry a negative value judgment. Moreover, there are many examples that conformity can lead to disaster (e.g., suicides, fraternity hazing, sexual assault). Nevertheless, conformity can also be used for many worthwhile purposes like stimulating pro-ecological or anti-racist behavior \cite{San:10}. Therefore, the question ``is conformity good or bad?'' has no scientific answer. Assuming the values most of us share, we can say that conformity is at times bad (e.g., when it leads someone to drive drunk), at times good (e.g., when it inhibits people from cutting into a theater line), and at times inconsequential (when it disposes tennis players to wear white) \cite{Mye:10}. Assumption that we make in this paper is that all these bad and good experiences related to the individual's behavior, conformist or independent, are collected in personal memory and influence somehow future behavior. So, if someone gained more benefits from being independent in the past, he will more likely behave independently in the future and vice versa. This means that we do not assume a priori that agents have personalities. At the beginning the system is homogeneous, i.e., all agents are the same. However, as the time goes, their individual situational experiences may lead to heterogeneity. 

To build the model based on the above assumption we need to know how person's memories of the past may influence their decisions about the future. The issue how memory impacts various decisions has been investigated in a number of psychological experiments related to political voting \cite{Mic:14}, evaluations of pleasurable experiences \cite{Do:Rup:Wol:08}, or episodes of pain \cite{Kah:Fre:Sch:Red:93,Red:Kat:Kah:03}.
For example, it has been shown empirically that patients' memories of the past may influence their decisions about the future medical treatment, yet memories are imperfect and susceptible to bias. In particular, the duration of the episode of pain has relatively little effect on subsequent evaluations, whereas the worst part of the experience and the amount of pain just before the episode ends are weighted heavily in the final impression. This observation, known presently under the name \textit{peak-end rule}, has been recently applied to a random walk model where the probability of moving left or right depends on the maximum value of a random variable associated with each time step \cite{Har:15}.
Here, we will use the same idea to model how agents decide to behave, independently or conform to social norms.

\section{Model description}
\label{sec:model}
We consider a system composed of $N$ mutually connected agents so that each agent is a neighbor of everyone else. Such a structure corresponds to a social network represented by a complete graph where each node is associated with one agent, and links indicate possible interactions between individuals. 
Networks like this one are frequently used to model relations in small groups or cliques where everybody knows each other \cite{Byr:etal:16, Kar:Sri:Cha:17}.
In order to mimic interactions in larger societies where the social structure is more complicated, complex networks are embraced, and they \textcolor{mycolor}{form} a framework for agent-based models \cite{Jed:17,Suc:Egu:Mig:05,Jav:Squ:15}.
An agent in $i$-th vertex is characterized by a two-state variable $s_i=\pm 1$, $i\in\{1,2,...,N\}$, which can be interpreted as a binary opinion on a given issue, e.g., an agent agrees or disagrees with something.
Of course, the interpretation depends strictly on a phenomenon that is modeled. 
Since the variable $s_i$ takes only two values just like a spin in the famous Ising model, agents are often called spins. Herein, we will use terms agent, spin, voter, individual interchangeably.
Agents in the system are subjected to two types of interactions that may change their opinions.
From a psychological point of view, these factors are recognized as two different responses to social pressure: conformity and independence \cite{Nai:Szn:16}.
The first one tends to order the system whereas the second one tries to disturb it and thus is regarded as stochastic noise.
In the model, a random sequential updating scheme is used. Therefore, at each elementary time step, we choose randomly one agent that can reconsider its opinion. Additionally, we choose at random a group of influence comprised of $q$ its nearest neighbors.
The group is called $q$-panel, and it attempts to impact upon a state of the chosen agent.
With probability $1-p$, the agent behaves as a conformist, and it yields to the group pressure by taking the same opinion of the panel each time when the group is unanimous -- all $q$ individuals are in the same state. Otherwise, when there is no consensus in the panel, nothing happens, and the state of the chosen \textcolor{mycolor}{agent} remains unchanged.
With complementary probability $p$, the agent acts independently of its neighbors.
It decides by its own whether to change its opinion, so with the probability $1/2$, the agent's state is changed to the opposite one.

In the previous studies, the parameter $p$, which controls the level of independence in the system, did not change in time. It was considered as an external parameter established at the beginning of the simulation so that all agents were equally likely to be independent in a given time step. Such an approach is called situation-oriented one because the \textcolor{mycolor}{agent}'s behavior (i.g., acting as a conformist or as an independent individual) may change from one time step to other during the simulation \cite{Nyc:Szn:13,Nyc:Szn:Cis:12}. Such a probabilistic situational approach has been used already earlier within the Galam's majority
model with contrarians\cite{Gal:04, Gal:07}.
From physical point of view, one can also think of these fluctuations in agents' attitude as the annealed disorder introduced to the system.
A competitive approach named person-oriented one has been studied, as well. It assumes that the behavior of agents arises from their personal traits, therefore, it does not evolve in time. So at the beginning of the simulation, a fraction $p$ of all individuals are set to always act independently, and the rest of agents are always conformists \cite{Szn:Szw:Wer:14, Jed:Szn:17}. 
This approach, on the other hand, can be related to the quenched disorder since the behavior of agents is frozen in time.
As it turned out, there are qualitative differences between these two approaches applied to the $q$-voter model with independence described above -- discontinuous phase transitions are present only in the situation-oriented model \cite{Jed:Szn:17}. Moreover, within the quenched approach the critical value of independence, below which an ordered phase with the majority-minority broken symmetry is observed, increases with the size of the influence group $q$. The same result has been obtained also for the $q$-voter model with anticonformity \cite{Nyc:Szn:Cis:12} and the Galam's majority model with contrarians \cite{Gal:04, Gal:07}.
In the annealed version of the $q$-voter model with independence the opposite relation is observed -- the critical value of
independence decreases with the size of the influence group $q$.

\begin{figure}[!t]
	\centerline{\epsfig{file=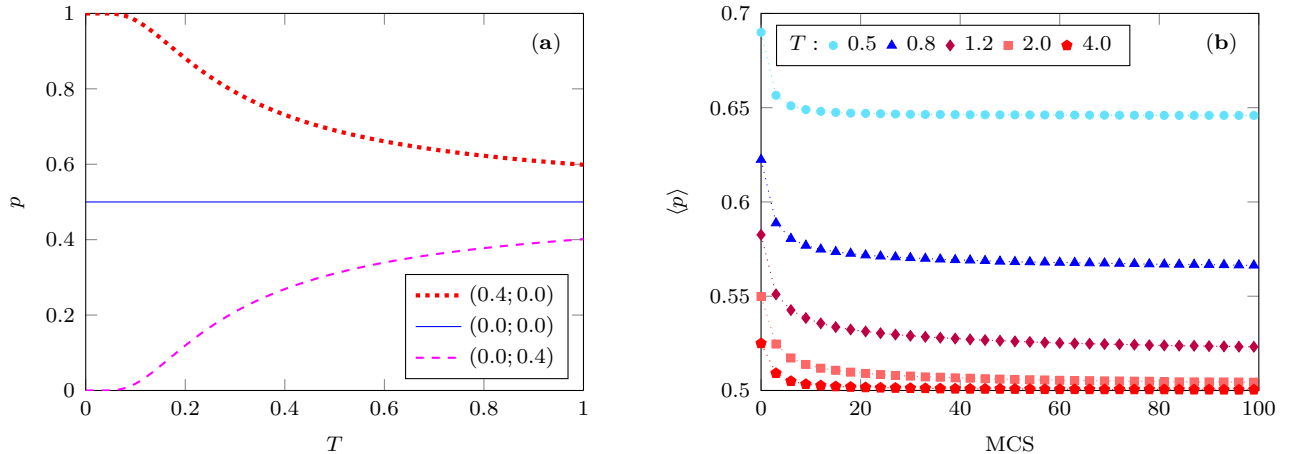}}
	\caption{\label{fig:independenceF} (a) Levels of independence $p$ as functions of the noise strength $T$ for different fixed values of the utilities \textcolor{mycolor}{$(u^I;u^C)$}. (b) Time evolution of the average levels of independence in the system of the size $N=10^6$. The initial utilities are as follows: \textcolor{mycolor}{$u^I_0=0.4$ and $u^C_0=0$}. Markers represent the outcome of Monte Carlo simulations and are connected just to guide the eye. Each line refers to a different value of $T$.  The noise increases from the top to the bottom. Note that the final values of $\langle p\rangle$ depend on the noise level, in contrast to the case with the same initial utilities where $\langle p\rangle=0.5$ independently of the noise.}
\end{figure}

In this work, on the other hand, $p$ is an internal parameter, which is not fixed and can alter in time depending on the past experiences of a given agent.
In particular, we incorporate the idea of the extreme value memory \cite{Har:15} into the $q$-voter model.
With each time step, we associate a random variable $U_t$ that represents  the utility, \textcolor{mycolor}{i.e.}, the amount of satisfaction that an agent receives from its choice.
We assume that these random variables $U_t$ are independent and identically distributed for different times.
Now, \textcolor{mycolor}{agents} remember the maximum value of $U_t$ for all their independent behaviors $U^I$ and separately for all their conformist behaviors $U^C$.
The next choice of their attitude is established based on these remembered maximum utilities in such a way that more probable is the behavior from which we had the most satisfaction in the past.
Therefore, \textcolor{mycolor}{we choose} the level of independence $p$ \textcolor{mycolor}{to be} given by a logistic function commonly used in discrete choice models in economics. For specific values of random variates $u^I$ and $u^C$, outcomes of random variables  $U^I$ and  $U^C$, the level of independence can be computed by the following formula:
\begin{equation}
\label{eq:independence}
p=\frac{e^{u^I/T}}{e^{u^I/T}+e^{u^C/T}},
\end{equation}
where $T$ is a positive external parameter, and it represents the noise level in the decision making process.
The maximum utilities $U^I$ and $U^C$ at the time $t$ are given by
\begin{align}
U^I &=\max_{0< \textcolor{mycolor}{j}< t}\{\textcolor{mycolor}{u^I_0, U_j}: \textrm{an agent is independent at \textcolor{mycolor}{$j$-th} step}\},\\
U^C &=\max_{0< \textcolor{mycolor}{j}< t}\{\textcolor{mycolor}{u^C_0, U_j}: \textrm{an agent is a conformist at \textcolor{mycolor}{$j$-th} step}\},
\end{align}
where \textcolor{mycolor}{$u^I_0$ and $u^C_0$} are the initial values of the utilities assigned to all agents. Note that when \textcolor{mycolor}{$u^I_0=u^C_0$}, the level of independence $p$ is equal to $1/2$ at the beginning of simulation, and initially, there is no bias in the attitude of agents, see Fig.~\ref{fig:independenceF}.
Additionally, the relation between $p$ and the noise level, that is, Eq.~(\ref{eq:independence}) is exemplified in Fig.~\ref{fig:independenceF} for three different pairs of utilities.
We should emphasize that $p$ not only changes in time, but also it differs from agent to agent since its value depends on the history of the specific individual. 

In general, $U_t$ can come from any distribution. However, throughout the work, we use an exponentially distributed random variable with parameter $\lambda>0$. Thus, the probability density function of $U_t$ has the following form:
\begin{equation}
\label{eq:pdfU}
	f_{U_t}(u)=\mathbbm{1}_{\{u\ge 0\}}\lambda e^{-\lambda u},
\end{equation}
where $\mathbbm{1}_{\{u\ge 0\}}$ is the indicator function defined on the interval $[0,\infty)$. 
This particular choice of a distribution is motivated by the emergence of an interesting transition between two stable states of the system \cite{Har:15}, which is described in the next section.

The algorithm of a simulation is following:
\begin{enumerate}
	\item Initialize the simulation ($t=0$): choose an initial concentration of up-spins $c(0)$, then assign an initial opinion to each agent -- for all vertices numbered by $i\in\{1,...,N\}$ draw a random number $r_i$ from a uniform distribution supported on the interval $[0,1]$.
	If $r_i<c(0)$, set $s_i(0)=1$. Otherwise, set $s_i(0)=-1$. Select any two real numbers as the initial values of utilities \textcolor{mycolor}{$u_0^I$ and $u_0^C$}. Note that all agents have the same initial values.
	\item Set $t=t+\Delta t$, and choose randomly $i$-th agent from the system.
	\item Draw two random numbers $r$ and \textcolor{mycolor}{$u_t$}, the first from the uniform distribution on $[0,1]$ and the second from the utility distribution given by Eq.~(\ref{eq:pdfU}).
	\item Calculate $p$ for the chosen agent based on its extreme value memory from Eq.~(\ref{eq:independence}).
	\item If $r<p$, the agent acts independently -- flip the spin with probability $1/2$. Relate \textcolor{mycolor}{$u_t$} with independence; go to 2.
	\item If $r\ge p$, the agent acts as a conformist -- choose randomly a group of $q$ distinct neighbors of the chosen $i$-th agent: $n_1, n_2,...,n_q\in\{1,2,...,N\}$. If all $q$ \textcolor{mycolor}{agents} are in the same state, the chosen agent takes the same opinion as the group. Relate \textcolor{mycolor}{$u_t$} with conformity; go to 2.
\end{enumerate}
As usual, one Monte Carlo step (MCS) corresponds to $N$ elementary time steps $\Delta t=1/N$.
\section{Results}
\label{sec:results}
In Ref.~\cite{Har:15}, a similar decision making mechanism was applied to a single, one-dimensional random walker that based on his past experiences had to choose the direction of its next step.
It turns out that the long-time walker's behavior is dependent on the utility distribution.
In particular, when $U_t$ comes from an exponential distribution with parameter $\lambda$, there is a transition at 
\begin{equation}
\label{eq:temperature}
T=T_c=\frac{1}{\lambda }
\end{equation}
\begin{figure}[!b]
	\centerline{\epsfig{file=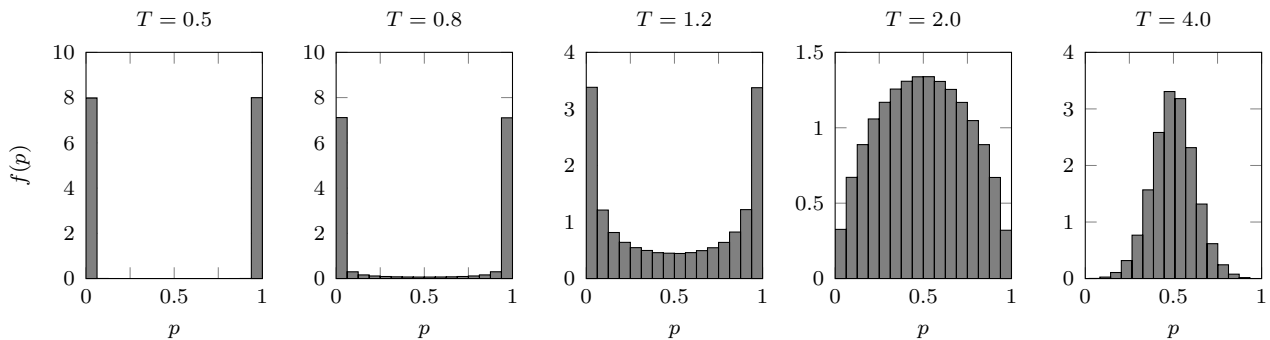}}
	\caption{\label{fig:histogramSym} Normalized histograms of the independence level $p$ in the system comprised of $N=10^6$ agents after $10^3$ MCS for the same initial utilities \textcolor{mycolor}{$u_0^I=u_0^C=0$} at five different noise levels increasing from the left to the right. The group of influence contains $q=4$ agents. Note that although the distribution changes with $T$, it stays symmetric all the time. Moreover, the average value of independence in the system $\langle p\rangle$ is not affected by the noise. It remains unchanged for different $T$ and equals $\langle p\rangle=0.5$.}
\end{figure}
\begin{figure}[!b]
	\centerline{\epsfig{file=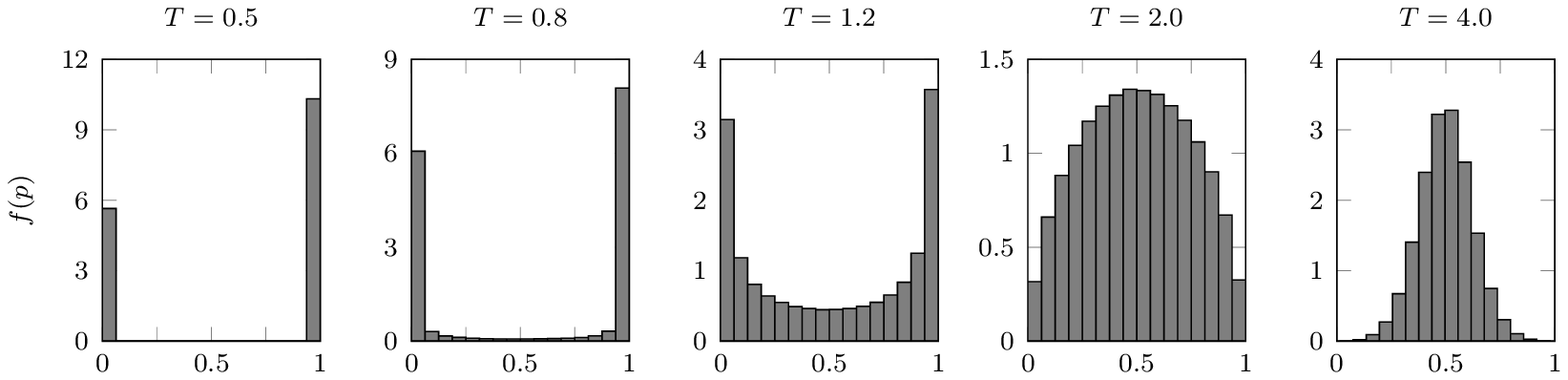}}
	\centerline{\epsfig{file=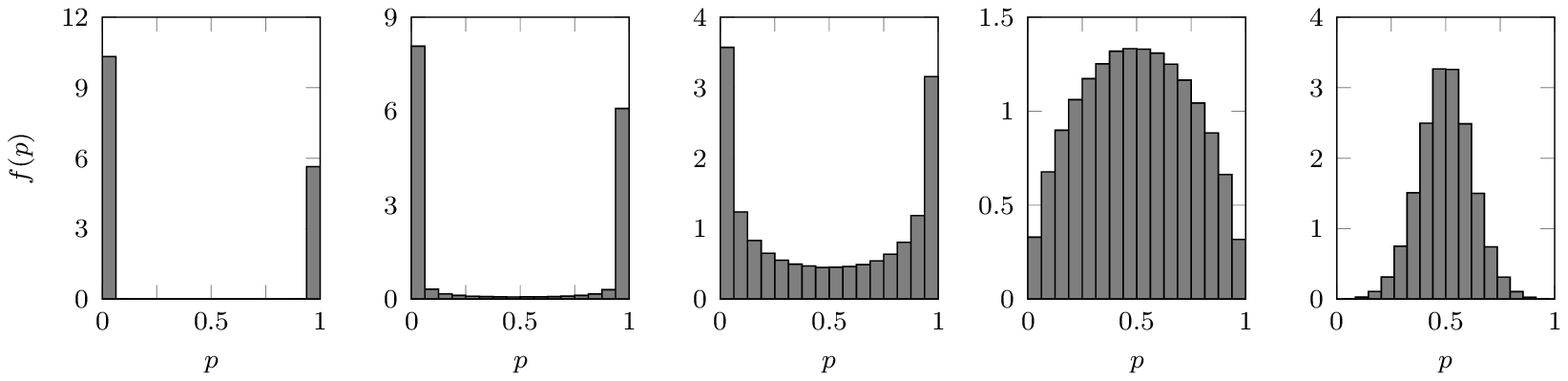}}
	\caption{\label{fig:histogramPanel} Normalized histograms of the independence level $p$ in the system comprised of $N=10^6$ agents with the bias in the initial values of utilities at five different noise levels increasing from the first column on the left to the last column on the right. The top row corresponds to the following initial values \textcolor{mycolor}{$u^I_0=0.4$ and $u^C_0=0$} while the bottom row refers to the values \textcolor{mycolor}{$u^I_0=0$ and $u^C_0=0.4$}. The data are collected after $10^3$ Monte Carlo steps.}
\end{figure}
between a stable state in which the walker is frozen into the motion along one direction, this happens for $\lambda T<1$, and a stable state in which it samples both directions roughly equally for $\lambda T>1$.
In terms of our model, the motion in one direction would correspond to an agent that does not change its behavior, i.e., it is permanently a conformist or an independent individual. On the other hand, the state with the equal sampling of both directions would refer to an agent that constantly changes its attitude from being independent to being a conformist. 
In fact, these states may be thought of as two separate regimes with quenched (person-oriented approach) and annealed disorders (situation-oriented approach) \cite{Jed:Szn:17,Szn:Szw:Wer:14}.
Now, the question arises whether the same transition will be observed in the system of interacting agents defined in the previous section, and how memory will impact the collective behavior of the system.
In order to bring answers to the above issues, we carry out Monte Carlo simulations of the $q$-voter model with independence and the extreme value memory where the utilities are drawn from the exponential distribution with $\lambda=1$. 
In our model, the level of independence $p$ is responsible for the behavior of agents.
Since $p$ is the property of every individual, we can examine the distribution of this feature in the system.
Figure~\ref{fig:histogramSym} presents the empirical density function \textcolor{mycolor}{$f(p)$} of the independence level after $10^3$ Monte Carlo steps for the system comprised of $N=10^6$ agents for several values of $T$.
In this case, the initial utility values are set to zero, i.e., \textcolor{mycolor}{$u_0^I=u_0^C=0$}.
Indeed, when $T$ is low, agents quickly promote one attitude over the other, and similarly as the walker became frozen into the motion along one direction, our agents become frozen into one type of behavior. Moreover, we can see from Fig.~\ref{fig:histogramSym} that approximately half of all agents are constantly independent, that is, their $p$ is close to 1, and the other half are conformists since their $p$ is close to 0.
On the other hand, when $T$ is high, the distribution of $p$ concentrates around the value $p=0.5$, so agents constantly change their behavior from one time step to the other.
On average, half of their choices involve conformity and the other half independence; see the right panel of Fig.~\ref{fig:histogramSym}.
Similar symmetric functions \textcolor{mycolor}{$f(p)$} are obtain for the initial utilities that are different from 0 but of the same value, i.e., \textcolor{mycolor}{$u_0^I=u_0^C$}.
In all these cases, the average value of independence $\langle p\rangle$ in the system is not influenced by the noise, and it remains at the same level $\langle p\rangle=0.5$ for different $T$.
In contrast, when \textcolor{mycolor}{$u_0^I\neq u_0^C$}, the distribution of the level of independence becomes skewed for low values of $T$, and only strong noise makes it symmetric, as we can see from Fig.~\ref{fig:histogramPanel}. 
Moreover, now the average value of independence in the system relies on $T$, see Fig.~\ref{fig:independenceF}.

\begin{figure}[!t]
	\centerline{\epsfig{file=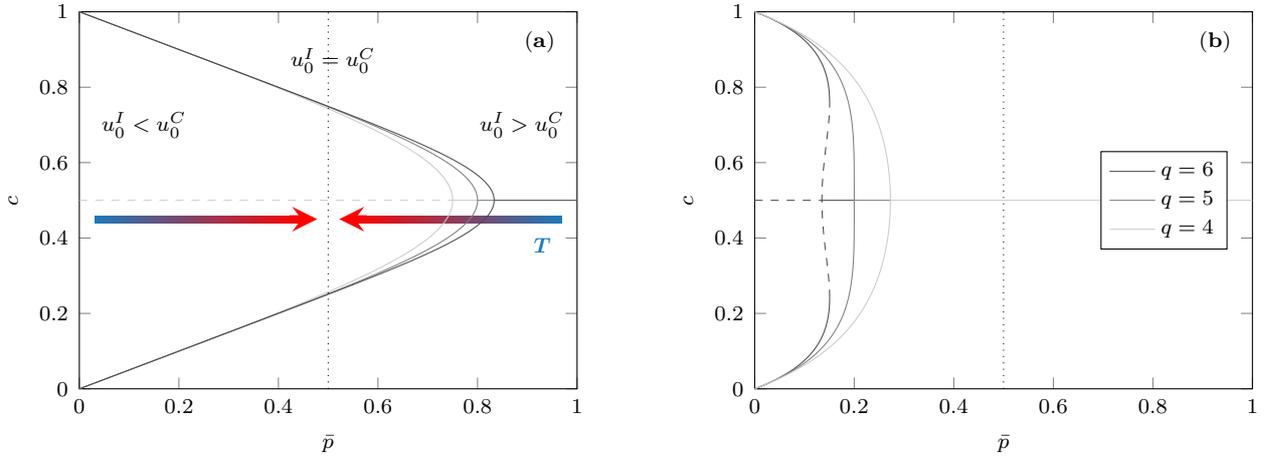}}
	\caption{\label{fig:phaseDiagramsQVoter} Phase diagrams for the $q$-voter model in the (a) quenched and (b) annealed approaches. Gray lines indicate the stationary values of the up-spin concentration given by Eqs.~(\ref{eq:qurnchedc}) and (\ref{eq:annealedc}), continuous and dashed ones refer to stable and unstable states accordingly. 
	(a) For the quenched case, when the initial utilities are equal, we have $\bar{p}=0.5$ (dotted vertical line) for all $T$. Lower values of the average level of independence are related to the situation in which \textcolor{mycolor}{$u_0^I<u_0^C$}. Higher values, on the other hand, correspond to the following utilities \textcolor{mycolor}{$u_0^I>u_0^C$}.  For these assymetric initial conditions, arrows illustrate how $\bar{p}$ will change when we increase the noise $T$ in the system.
	(b) For the annealed case, the average level of independence in the system amounts to $\bar{p}=0.5$ (dotted vertical line), and the only stable value of the stationary up-spin concentration is $c=0.5$.
	Note that for the annealed case, $\bar{p}$ corresponds to the probability of an agent being independent in a given time step whereas for the quenched one, $\bar{p}$ refers to the fraction of agents in the system that are independent all the time.}
\end{figure}

In general, we can obtain diverse stationary probability density functions of the level of independence. The shape of them depends on the initial values of utilities and the level of noise $T$.
Although we are not able to derive strictly the formula for \textcolor{mycolor}{$f(p)$}, let us assume that we already know the form of this density function.
In that case, we can calculate the stationary value of the up-spin concentration in the system defined as a fraction of agents with a positive opinion, i.e., $s_i=1$. If there are $N_\uparrow$ \textcolor{mycolor}{agents} with $s_i=1$ in a given time step, the up-spin concentration is as follows
\begin{equation}
	c=\frac{N_\uparrow}{N}.
\end{equation}
In every elementary time step, the total number of agents with a positive opinion may increase by one, decrease by one, or remain at the same level since we use a random sequential updating. This corresponds to the following transition rates:
\begin{equation}
\begin{split}
\gamma^+ &=\text{P}\left(c\rightarrow c+\frac{1}{N}\right),\\
\gamma^- &=\text{P}\left(c\rightarrow c-\frac{1}{N}\right),\\
\end{split}
\end{equation}
so that the time evolution of the concentration is given by the rate equation
\begin{equation}
	\label{eq:rateEq}
	\frac{\partial c}{\partial t}=\gamma^+-\gamma^-,
\end{equation}
in the limit of $N\rightarrow\infty$. Note that for the stationary value of the up-spin concentration, Eq.~(\ref{eq:rateEq}) gives $\gamma^+=\gamma^-$.
Similarly, in the quenched region, the concentration of up-spins only among agents with the independence level in the range $(p,p+dp)$ can be introduced. We use the following notation $c_p$ to denote such defined quantity\textcolor{mycolor}{:
\begin{equation}
	c_p=\frac{N^\uparrow_p}{N_p},
\end{equation}
where $N_p$ is the number of agents with the specific independence level, and $N_p^\uparrow$ indicates how many of them have a positive opinion.}
Now, thanks to the above partition, we can write down explicitly the transition rates $\gamma_p^+$, $\gamma_p^-$ for all values of independence, and consequently, find the stationary levels of the concentrations $c_p$.
Based on the model description and the mean-field approximation \cite{Mor:etal:13,Nyc:Szn:13}, the transition rates have the following forms:
\begin{equation}
\begin{split}
\gamma^+_p&=(1-c_p)\left[(1-p)c^q+\frac{p}{2}\right],\\
\gamma^-_p&=c_p\left[(1-p)(1-c)^q+\frac{p}{2}\right].
\end{split}
\end{equation}
By equalizing the above equations, we obtain the stationary concentration of agents with a positive opinion and with the level of independence close to $p$:
\begin{equation}
\label{eq:cp}
c_p=\frac{(1-p)c^q+\frac{p}{2}}{(1-p)\left[c^q+(1-c)^q\right]+p}.
\end{equation}
Having $c_p$, we are able to determine the overall up-spin concentration in the quenched case by solving the following self-consistent equation
\begin{equation}
\label{eq:selfConsistent}
c=\int_{-\infty}^\infty c_pf(p)dp.
\end{equation}
In the quenched region, since the probability density function \textcolor{mycolor}{$f(p)$} concentrates around points 0 and 1, as seen in Figs.~\ref{fig:histogramSym} and \ref{fig:histogramPanel}, it can be approximated by a discrete distribution represented as a sum of two Dirac delta functions at these points with certain masses $\bar{p}$ and $1-\bar{p}$ accordingly:
\begin{equation}
\label{eq:quenchedDistribution}
f(p)=\bar{p}\delta(p-1)+(1-\bar{p})\delta(p),
\end{equation}
where $\bar{p}$ corresponds to the average level of independence in the system because
\begin{equation}
	\langle p\rangle=\int_{-\infty}^\infty pf(p)dp=\bar{p}.
\end{equation}
Note that $\bar{p}$ can be also interpreted as a fraction of independent individuals in the quenched system. For $\bar{p}=0.5$, we get the symmetric distribution which is obtain when the initial utilities are equal \textcolor{mycolor}{$u_0^I=u_0^C$}. In contrast, if at the beginning the utilities are not equal and \textcolor{mycolor}{$u_0^I<u_0^C$}, then we have $\bar{p}<0.5$. The opposite situation takes place for \textcolor{mycolor}{$u_0^I>u_0^C$}; see the first column of Fig.~\ref{fig:histogramPanel}.
Now, let us perform the integration in Eq.~(\ref{eq:selfConsistent}) after inserting Eqs.~(\ref{eq:cp}) and (\ref{eq:quenchedDistribution}) into it. The following implicit expression is obtained for the stationary up-spin concentration at low noise values $T$ where we can approximate \textcolor{mycolor}{$f(p)$} by two Dirac delta functions Eq.~(\ref{eq:quenchedDistribution}):
\begin{equation}
\label{eq:qurnchedc}
c=\frac{1}{2}\bar{p}+\frac{c^q}{c^q+(1-c)^q}(1-\bar{p}).
\end{equation}
Figure~\ref{fig:phaseDiagramsQVoter} illustrates graphically the above dependency for a few values of $q$. 
Continuous and dashed lines refer to stable and unstable states accordingly.
Now, if we knew the connection between the average independence $\bar{p}$, the initial utilities \textcolor{mycolor}{$u_0^I$ and $u_0^C$}, and the noise level $T$, we would read from Fig.~\ref{fig:phaseDiagramsQVoter} or directly from Eq.~(\ref{eq:qurnchedc}) the stationary value of the up-spin concentration for these parameters in the quenched regime, i.e., for small $T$.
In general, this relation is unidentified, however, we already know that $\langle p\rangle=0.5$ for all values of $T$ for the special case when \textcolor{mycolor}{$u_0^I=u_0^C$}.
Thus, when the initial utilities are the same, in order to establish the stationary concentrations for small noises, we look at Fig.~\ref{fig:phaseDiagramsQVoter} and the intersections of solid lines representing the stable solutions of Eq.~(\ref{eq:qurnchedc}) with the dotted vertical line at $\bar{p}=0.5$.
We see that there are only two such intersections. Moreover, we expect that $q$ should not influence much the level of the stationary up-spin concentration in the quenched area since changing $q$ does not shift a lot the intersection points.

In the asymmetric case, that is to say, when \textcolor{mycolor}{$u_0^I\neq u_0^C$}, the situation is different. As seen in Fig.~\ref{fig:histogramPanel}, we start from the skewed distribution, i.e., $\bar{p}\neq 0.5$ for small $T$, but along with the increasing level of noise, \textcolor{mycolor}{$f(p)$} becomes more symmetric, and $\bar{p}$ approaches 0.5. 
Of course, this happens only to some extent since for higher $T$, Eq.~(\ref{eq:quenchedDistribution}) does not approximate well the real distribution.
The direction from which we approach this point depends on the initial utilities and is illustrated by an arrow in Fig.~\ref{fig:phaseDiagramsQVoter}. If \textcolor{mycolor}{$u_0^I<u_0^C$}, the average value of independence will increase to $\bar{p}= 0.5$.
On the other hand, when \textcolor{mycolor}{$u_0^I>u_0^C$}, it will decrease to that value.
Therefore, raising the level of noise should decrease the order for \textcolor{mycolor}{$u_0^I<u_0^C$}, that it, the stationary values of the up-spin concentration tend towards a disordered phase $c=0.5$, and at the same time, increase it for \textcolor{mycolor}{$u_0^I>u_0^C$}; see Fig.~\ref{fig:phaseDiagramsQVoter}.
Particularly interesting is the second case because greater noise boosts strength of ordering in the system, which is untypical.

In the annealed region, the probability density function \textcolor{mycolor}{$f(p)$} concentrates around one point $\bar{p}$; see Figs.~\ref{fig:histogramSym} and \ref{fig:histogramPanel}. Therefore, it can be approximated by a single Dirac delta function at this point:
\begin{equation}
\label{eq:annaledDistribution}
f(p)=\delta(p-\bar{p}).
\end{equation}
Once again $\bar{p}$ refers to the average level of independence in the system. We know that in our case $\bar{p}=0.5$, but we can consider more general one.
Such annealed version of the $q$-voter model was already considered in \cite{Nyc:Szn:Cis:12}, and the stationary
up-spin concentration has the following form in this case:
\begin{equation}
\label{eq:annealedc}
c=\frac{(1-\bar{p})c^q+\frac{\bar{p}}{2}}{(1-\bar{p})\left[c^q+(1-c)^q\right]+\bar{p}}.
\end{equation}
We present the above dependency in Fig.~\ref{fig:phaseDiagramsQVoter} for different values of $q$.
As seen, the only stationary value of the up-spin concentration that can be attained for $\bar{p}=0.5$ is $c=0.5$.
It means that for large values of $T$, in the annealed regime, our system is disordered.

\begin{figure}[!t]
	\centerline{\epsfig{file=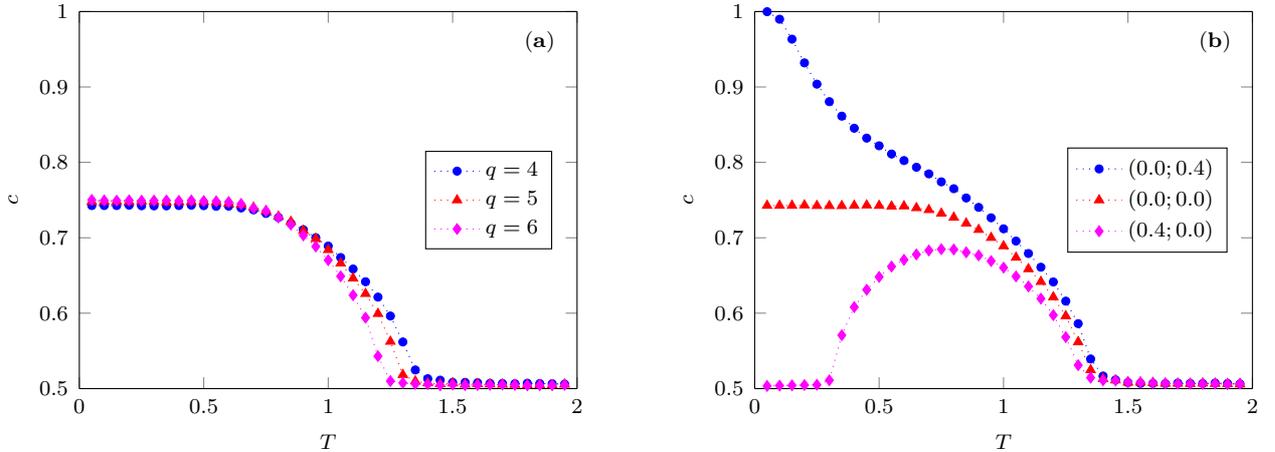}}
	\caption{\label{fig:phaseDiagrams} Phase diagrams for the $q$-voter model with extreme value memory and exponentially distributed utilities with parameter $\lambda=1$. \textcolor{mycolor}{Initially, all spins are up, so $c(0)=1$.} Dots represent the outcome of Monte Carlo simulations of the system containing $N=10^4$ agents. The results are averaged over $10^2$ runs and collected after $10^3$ MCS. Dotted lines are drawn just to guide the eye. (a) The model with no bias in the initial values of the utilities, i.e., \textcolor{mycolor}{$u_0^I=u_0^C=0$} for different $q$-panel sizes. (b) The model with the group of influence comprised of $q=4$ agents and different initial conditions for the utilities \textcolor{mycolor}{$(u_0^I; u_0^C)$}.}
\end{figure}

Our theoretical predictions agree with the Monte Carlo simulations of the system. The stationary value of the up-spin concentration as a function of the noise level $T$ is presented in Fig.~\ref{fig:phaseDiagrams} \textcolor{mycolor}{ for the initial value of the up-spin concentration $c(0)=1$}. Phase diagrams in the left panel correspond to the model with the same initial values of both utilities \textcolor{mycolor}{$u_0^I=u_0^C=0$} but different sizes of the group of influence. In the right panel, on the other hand, we see how the results change if we introduce a bias in these initial conditions for the fixed group size $q=4$.
For the cases where \textcolor{mycolor}{$u_0^I=u_0^C$}, a transition between an ordered phase at low noise levels and a disordered phase at high noise levels is observed.
Moreover, the number of agents in the $q$-panel does not affect much the up-spin concentration in the quenched regime, i.e., in the region where the noise $T$ is sufficiently small. 
When \textcolor{mycolor}{$u_0^I\neq u_0^C$} depending on which utility is \textcolor{mycolor}{greater}, we can control the value of the up-spin concentration for small $T$. 
For \textcolor{mycolor}{$u_0^I< u_0^C$}, the concentration of agents with a positive opinion decreases along with the increasing noise. For the opposite case, when \textcolor{mycolor}{$u_0^I> u_0^C$}, we have very different situation. Initially, the concentration unusually rises with the rising level of noise but only to some point after which it starts decreasing. As predicted, the first phase is connected with the quenched regime where bigger noise makes the bimodal distribution of $p$ more symmetric, then $c$ rises. Nonetheless, for large enough $T$ the distribution eventually changes its shape to unimodal in the annealed area, and $c$ decreases.


\section{Conclusions}
\label{sec:conclusion}
Which of these two approaches, situation-oriented or person-oriented, is more reasonable? The answer for this question was the subject of a long-term person-situation debate \cite{Don:Luc:Fle:09}. Particularly illustrative metaphor related to this issue has been given by one of the most creative social psychologists Richard E. Nisbett, who claims that behavior can be predicted much better from the social setting than from personality traits \cite{Nis:80}. However, the metaphor that he used, as well as modern personality psychology suggest that both approaches are reasonable depending on the investigated problem. Let us quote here Nisbett's metaphor that was aimed to convince people why believing that personal traits can predict individual's behavior is naive and reminds ancient beliefs about physical world: ``\textit{(...) in ancient physics, the behavior of objects was understood exclusively in terms of the object itself: A stone sinks when placed in water because it has the property of ``gravity''; a piece of woods floats because it has the property of ``levity''. In modern physics, in contrast, an understanding of the behavior of objects requires simultaneous specification of the environmental forces, the properties of the object, and the relation between the environmental forces and the properties of the object.}''. As physicists we know that a stone and a piece of wood indeed fall differently in water because of their individual traits, such as density of the material and the size of an object. However, in the air these differences will be less visible, and in vacuum both objects will fall identically. Such a metaphor may seem  too far reaching, yet social psychologists have shown that often situation may completely overcome personality \cite{Mye:10,San:10,Nis:80}. However, it all depends on the external conditions.

In this paper we have shown that indeed, depending on the external factor related to social temperature $T$, we can observe one of two types of behavior: person or situation-oriented. We have developed the $q$-voter model with independence incorporating the idea of memory in the spirit of extreme value memory \cite{Har:15}. Initially the system is homogeneous, which means that  all agents have the same tendencies \textcolor{mycolor}{$u_0^I$} for being independent and \textcolor{mycolor}{$u_0^C$} for being conformists. However, due to the memories of the past experiences related to each type of social response they may acquire personal traits if the level of social temperature is below a critical value. We may relate the ratio between \textcolor{mycolor}{$u_0^I$ and $u_0^C$} to one of the most important Hofstede's cultural dimensions\cite{Hof:Hof:Min:10}, namely individualism vs. collectivism (IDV). Particularly interesting relation between public opinion and social temperature $T$ has been found for the initially individualistic societies \textcolor{mycolor}{($u_0^I>u_0^C$)}. It occurs that in such a society there is an optimal $T$ at which the agreement in the society is the highest. For low $T$, there is complete disagreement, i.e., the number of positive and negative opinions is the same ($c=1/2$). Above certain threshold a majority opinion appears $c\neq1/2$ and above the critical social temperature $T_c$ it again decreases to $c=1/2$ (stale-mate situation). For initially collective societies \textcolor{mycolor}{($u_0^I<u_0^C$)} \textcolor{mycolor}{the agreement in the society} decreases monotonically and reaches stale-mate situation above the critical social temperature $T_c$.

Spontaneous appearance of heterogeneity related to different types of social response, rational and inflexible, has been observed already in \cite{Mar:Gal:13}. They have combined the Galam unifying frame (GUF) \cite{Gal:05} with the CODA formalism proposed by Martins \cite{Mar:08}, in which each agent is described by two variables: continuous private opinion and related discrete choice. In CODA model an individual's  continuous opinion was used to measure how certain each agent was about its decision, and therefore, inflexibility could naturally occurred as a consequence of very strong private opinion. In our model agents are described by a single binary variable, and the heterogeneity appears as a consequence of past experiences related to the type of behavior, analogously as in \cite{Har:15}.

The extreme value memory is certainly not the only possibility to incorporate the idea of memory into the $q$-voter model although particularly interesting because of the empirical justification laying below the assumption as well as interesting outcome, which can be also nicely interpreted in terms of social psychology. However, it would be worth checking out how other types of memory would influence the opinion formation under the $q$-voter model, and we believe that this is an interesting task for the future studies.

\section*{Acknowledgements}
This work was supported by funds from the National Science Center (NCN, Poland) through grants no. 2016/23/N/ST2/00729 and no. 2016/21/B/HS6/01256.

\section*{References}
\bibliographystyle{elsarticle-num} 
\bibliography{mybib}





\end{document}